\documentclass[aps,prd,showpacs,superscriptaddress,groupedaddres]{revtex4}
\usepackage[latin1]{inputenc}

\usepackage{graphicx}
\usepackage{amsfonts}
\usepackage{amssymb}
\usepackage{color}
\usepackage{amsmath}
\usepackage{breqn}
\usepackage{comment}
\usepackage[colorlinks,linkcolor=blue,citecolor=blue]{hyperref}
\usepackage{aas_macros}
\usepackage{float}
\usepackage{babel}
\usepackage[font=small,labelfont=bf]{caption}
\usepackage{xspace}
\newcommand{\comm}[1]{}

\begin{document}
\input{epsf.sty}

\title{Cosmological Evolution in Bimetric Gravity: Observational Constraints and LSS Signatures}

\author{Ajay Bassi}
\email {ajay@ctp-jamia.res.in}
\affiliation{Centre for Theoretical Physics$, ~Jamia~Millia~Islamia,~ New~Delhi-110025,~ India.$}

\author{Shahnawaz A. Adil}
\email {shazadil14@gmail.com}
\affiliation{Department of Physics$, ~Jamia~Millia~Islamia,~ New~Delhi-110025, ~India.$}

\author{Manvendra Pratap Rajvanshi}
\email {manvendra155@gmail.com}
\affiliation{Centre for Theoretical Physics$, ~Jamia~Millia~Islamia,~ New~Delhi-110025,~ India.$}

\author{Anjan A. Sen}
\email{aasen@jmi.ac.in}
\affiliation{Centre for Theoretical Physics$, ~Jamia~Millia~Islamia,~ New~Delhi-110025,~ India.$}

\begin{abstract}

\noindent
Bimetric gravity is an interesting alternative to standard GR given its potential to provide a concrete theoretical framework for a ghost-free massive gravity theory. Here we investigate a class of Bimetric gravity models for their cosmological implications. We study the background expansion as well as the growth of matter perturbations at linear and second order. We use low-redshift observations from SnIa (Pantheon+ and SH0ES), Baryon Acoustic Oscillations (BAO), the growth ($f\sigma_{8}$) measurements and the measurement from Megamaser Cosmology Project to constrain the Bimetric model. We find that the Bimetric models are consistent with the present data alongside the $\Lambda$CDM model. We reconstructed the `` effective dark energy equation of state" ($\omega_{de}$) and "Skewness" ($S_{3}$) parameters for the Bimetric model from the observational constraints and show that the current low-redshift data allow significant deviations in $\omega_{de}$ and $S_{3}$ parameters with respect to the $\Lambda$CDM behaviour. We also look at the ISW effect via galaxy-temperature correlations and find that the best fit Bimetric model behaves similarly to $\Lambda$CDM in this regard.
\end{abstract}

\keywords{cosmology, modified gravity, Bimetric gravity, ISW effect, skewness, perturbations}

\maketitle

\section{Introduction}

Observed late time acceleration\cite{Riess_1998,Perlmutter_1999,1998ApJ...507...46S} of the average expansion of the Universe has been one of the central challenges of modern Cosmology for more than two decades. The two standard approaches to explain such late time acceleration have been: 1) assuming the presence of a new unobserved component, called dark energy in the energy budget of the universe \cite{2010deto.book.....A} (similar to dark matter but with negative pressure instead of zero pressure) 2) modifying the standard theory of Gravitation (Einstein's General Relativity) at large cosmological scales \cite{2012PhR...513....1C,Nojiri_2011}. The concordance $\Lambda CDM$ model \cite{2021arXiv210505208P} has been tremendously successful in explaining observations across scales \cite{planck_2015_cosmo,planck_2018_cosmo}, despite a number of theoretical problems to construct such model \cite{2000astro.ph..5265W}. But recently for the first time, $\Lambda$CDM model is facing some serious questions related to observational results.
There is a discrepancy in values of Hubble constant ($H_0$) inferred from nearby cosmological observations \cite{SH0ES_2016ApJ...826...56R}, and those inferred from CMB observations \cite{planck_2015_cosmo,planck_2018_cosmo}. This tension, termed $H_0$ tension, has now reached at the level of $5\sigma$ and systematical errors as source for this tensions may not be enough to explain this tension \cite{Ins1_2018MNRAS.477.4534F,Ins2_2018A&A...609A..72D}. 

This has been the main driving force in the renewed exploration of the theoretical regime with dark energy models beyond $\Lambda CDM$  or  modified theories of gravity. In relation to modified theories of gravity, massive gravity models are one of the most studied modified gravity models. In standard theory of gravity, the gravity is described by massless spin-2 fields called gravitons which are still not observed. Fierz and Pauli \cite{1939RSPSA.173..211F} made the first attempt to make these spin-2 field massive with a linear theory. But it was later shown by Boulware and Deser \cite{1972PhRvD...6.3368B} that the nonlinear extensions of such theories contain a ghost (Boulware-Deser (BW) Ghost). A ghost-free, nonlinear theory for massive spin-2 field in flat space time  was first proposed by de Rham, Gabadadze and Trolley in 2011 \cite{2011PhRvL.106w1101D}, which is called the dRGT theory. Hassan and Rosen \cite{2012JHEP...02..126H} later extended the dRGT theory for Bimetric space time that describes a gravitating massive spin-2 field. Cosmological solutions in Bimetric gravity have been studied which can result in the late time acceleration of the Universe without any presence of explicit dark energy term. Different families of Bimetric gravity models have been studied \cite{Strauss_2012,2018JCAP...09..025M,Dhawan_2017,L_ben_2020,H_g_s_2021,H_g_s_2021a} in light of CMB, BAO and Supernovae data. Gravitational waves in these theories with some constraints on parameter space are studied in \cite{Bi_GW_2017PhRvL.119k1101M,Bi_GW_2017PhRvD..96b3518B}. Big Bang Nucleosynthesis constraints were obtained in \cite{Bi_BBN_2021arXiv210609030H}. 

On the other hand, theories with phantom behaviour(i.e. equation of state $w$ crossing below $-1$) \cite{2021Entrp..23..404D,Chudaykin_2022arXiv220303666C} have shown promise in reducing/overcoming the cosmological tensions, more specifically the Hubble tension although the theoretical underpinnings of phantom phenomenology are still under investigations as many of such theories may suffer from instabilities. The viable Bimetric gravity can also show phantom behaviour in some of their parameter space. Providing an effective $\Lambda$-like effect and the possibility of phantom behaviour can make Bimetric gravity suitable for investigation as a possible solution for the Hubble tension.

In the literature, there are not many studies related to the effect of Bimetric gravity on the large-scale structure formation in the universe. In particular, the effects of modification of gravity at cosmological scales in higher-order clustering are still mostly unknown. One of the important parameters related to higher order clustering is the ``skewness'' parameter \cite{Amendola_2004}. This is defined as the ``normalized third order moment in count-in-cells statistics'' which can describe the non-Gaussian feature in the probability distribution of the perturbed matter field. In a purely matter-dominated Einstein-de Sitter Universe, one gets $S_{3} \approx 34/7 \approx 4.857$. Any observed deviation from this value can be a signature for a modified gravity model \cite{2022arXiv220208355E}.

Our aim in this work is to study the linear and second-order perturbations for a specific class of Bimetric gravity models and illustrate the effects of these modifications to gravity on observables related to perturbed Universe such as $f\sigma_8$, skewness parameter $S_{3}$ and Integrated Sachs-Wolfe(ISW) effect and compare the results with concordance $\Lambda$CDM models. This can be particularly interesting in the context of results from future surveys like Euclid which can provide an accurate measurement of the skewness parameter and can potentially distinguish $\Lambda$CDM model from different modified gravity models including the Bimetric gravity.

This paper is organized as follows: in section \ref{sec:bi}, we briefly describe the theory of Bimetric gravity; in section \ref{sec:grow_lss}, we describe the perturbation theory formalism used for the Bimetric gravity and the discuss the effects of these perturbations on observables in respective subsections \ref{subsec:observable}. Data constraints are obtained in section \ref{subsubsec:DataA}. In section \ref{subsubsec:sw}, we study ISW effect through galaxy-temperature cross-correlations. We conclude with a discussion in the last section.

\section{Bimetric Gravity \& Cosmology}
\label{sec:bi}
Bimetric gravity is the theory of two interacting spin-2 fields, one massive and one massless. An interacting symmetric spin-2 field $f_{\mu\nu}$ is introduced together with the physical metric $g_{\mu\nu}$, which is a massless spin-2 field. The standard matter particles and fields are coupled to the physical metric $g_{\mu\nu}$ only. For ghost-free Bimetric gravity, where both metrics are dynamical, action can be written as \cite{2012JHEP...02..126H,2018JCAP...09..025M}
\begin{equation}
    S = -\int d^4x\sqrt{-g} \frac{R}{2m_g} - \int d^4x\sqrt{-f} \frac{\tilde{R}}{2m_f} 
        + m^4 V(g_{\mu\nu},f_{\mu\nu}) + \int d^4x\sqrt{-g}  L_m,   
\end{equation}
where $R$ and $\tilde{R}$ are Ricci scalars for $g_{\mu\nu}$ and $f_{\mu\nu}$. $m_g$ and $m_f$ are the Planck's mass for the  metrics $g_{\mu\nu}$ and $f_{\mu\nu}$ respectively.. $V$ is the interaction term which has the parametric form as
\begin{equation}
    V = \sum_{n=0}^4\beta_{n}e_n(\chi),
\end{equation}
where 
\begin{equation}
    \chi = \sqrt{g^{-1}f}.
\end{equation}

Here $e_n(\chi)$ is the elementary symmetry polynomials of eigenvalues of the matrix $\chi$ which can be written as follows 
\begin{equation}\label{e_n}
    \begin{aligned}
    e_0(\chi)&=1,\quad e_1(\chi)=[\chi],\quad e_2(\chi)=\frac{1}{2}\left([\chi]^2-[\chi^2]\right)\\
    e_0(\chi)&=\frac{1}{6}\left([\chi]^3-3[\chi][\chi^2]+2[\chi^3]\right), \quad e_4(\chi)=\textrm{det}(\chi)
    \end{aligned}
\end{equation}
whrere, $[\chi]$ is the trace of the matrix $\chi$ and $\textrm{det}(\chi)$ is the determinant of $\chi$. The Bimetric gravity is characterized by 5 constants $\beta_i$. 
Under the scaling transformation $f_{\mu\nu}\to \frac{m_g^2}{m_f^2}f_{\mu\nu}$ and $\beta_n\to \left(\frac{m_f}{m_g}\right)^n \beta_n$, we can make $M_{*}^2=1$ where, $M_{*}=m_f/m_g$. Hence $M_{*}$ is not a free parameter. In what follows, we consider $M_{*}^2=1$ and $m_g=m_{f}$.
With such rescaling \cite{2020JCAP...09..024L}, it is common to work in terms of dimensionless quantities which remain same under these scaling. Here we use the definitions and parameterisation of Dhawan et. al \cite{2018JCAP...09..025M} to write the dimensionless parameters as
\begin{equation}
    B_i\equiv \frac{m^2 \beta_i}{H_0}.
\end{equation}
There are various possible models  depending on which $Bs$ are non-zero. For simplicity, here we consider models with only $B_0$ and $B_1$ nonzero. Cosmological expansion for these models is given by \cite{2020ApJ...894...54D}

\begin{equation}
    \frac{H^2}{H_0^2} = \frac{\Omega_m(1+z)^3}{2} +\frac{B_0}{6} +
            \sqrt{\left( \frac{\Omega_m(1+z)^3}{2} +\frac{B_0}{6}  \right)^2 
                +\frac{B_1^2}{3}},
                \label{eq:Bimetric_H}
\end{equation}
where due to spatial flatness condition, the parameter $B_{0}$ can be written in terms of other two parameters $\Omega_{m}$ and $B_{1}$ as
\begin{equation}
    B_0 = 3\left( 1 - \Omega_m - \frac{B_1^2}{3}  \right),
\end{equation}
where $\Omega_{m}$ is the present day matter density parameter. As one can see in equation \ref{eq:Bimetric_H}, the model naturally gives a cosmological constant term $\frac{B_0}{3}$ in the Universe which can be positive or negative depending on the values of $\Omega_{m}$ and $B_{1}$. This originates from the interacting potential itself. For early times $z\to\infty$,
\begin{equation}
     \frac{H^2}{H_0^2} \approx{\Omega_m(1+z)^3} +\frac{B_0}{3},
\end{equation}
which is $\Lambda CDM$ model. For future infinity $(z\rightarrow -1)$,
\begin{equation}
     \frac{H^2}{H_0^2} \approx  constant,
\end{equation}
and hence a de-Sitter or anti-de-Sitter model.

Using the above equations and comparing them with the standard model, we can derive an expression for the effective dark energy equation of state ($w_{de}$). We plot $w_{de}(z)$ as a function of redshift($z$) in figure \ref{fig:wde} and observe that the Bimetric gravity can show phantom behavior. In the figure \ref{fig:wde_contour}, where we show the dependence of $w_{de}(z=0)$ on $B_{1}$ and $\Omega_{m0}$, it is evident that Bimetric gravity shows phantom behavior at present for a large range of parameter values and can provide a possible theoretical basis for parametric phantom models, which people have recently considered \cite{2021Entrp..23..404D, 2022MNRAS.tmp.2681S} to alleviate the cosmological tensions.

\begin{minipage}{0.975\textwidth}
\centering
 
  \begin{minipage}[b]{0.48\textwidth}
    \centering
   \includegraphics[width=\textwidth]{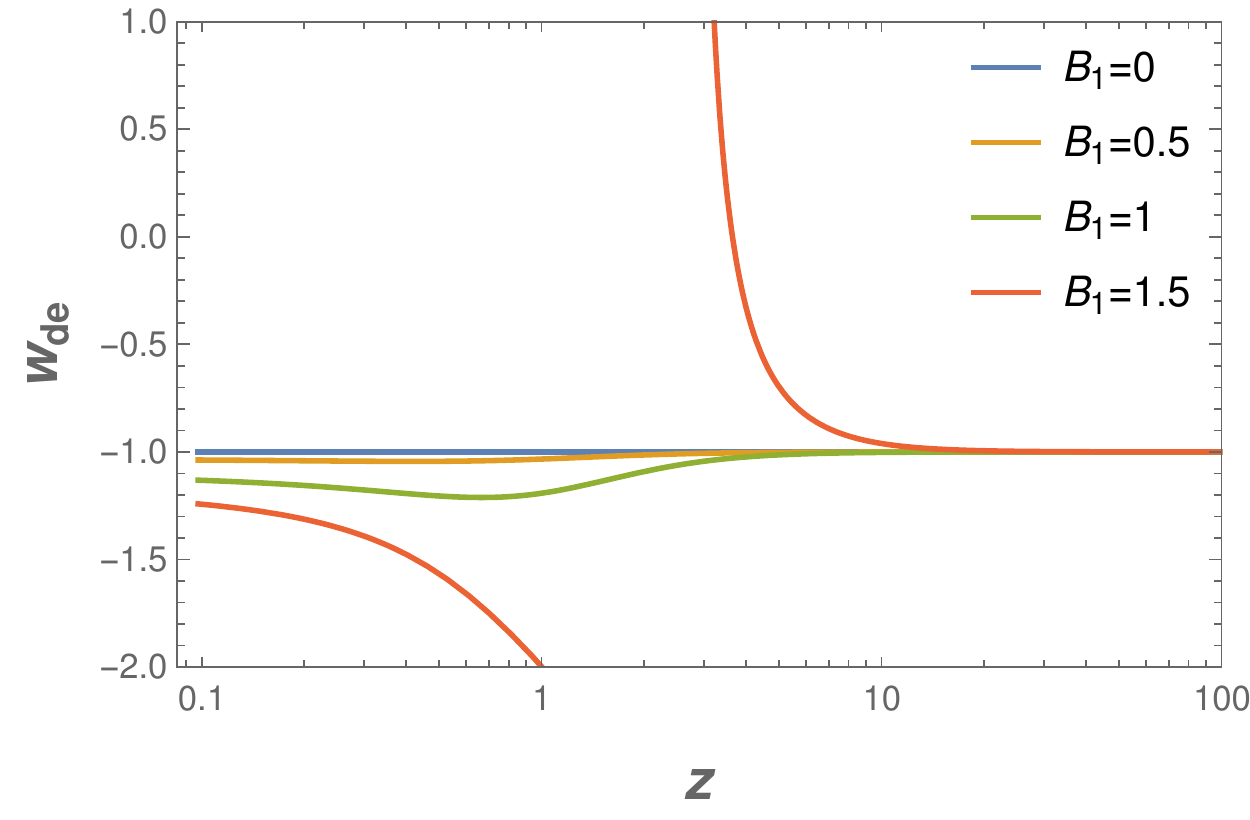}
  \captionof{figure}{Equation of state $w_{de}$ for different values of parameter $B_1$.} 
  \label{fig:wde}
    \end{minipage}
     \hfill
     \begin{minipage}[b]{0.48\textwidth}
    \centering
     \includegraphics[width=\textwidth]{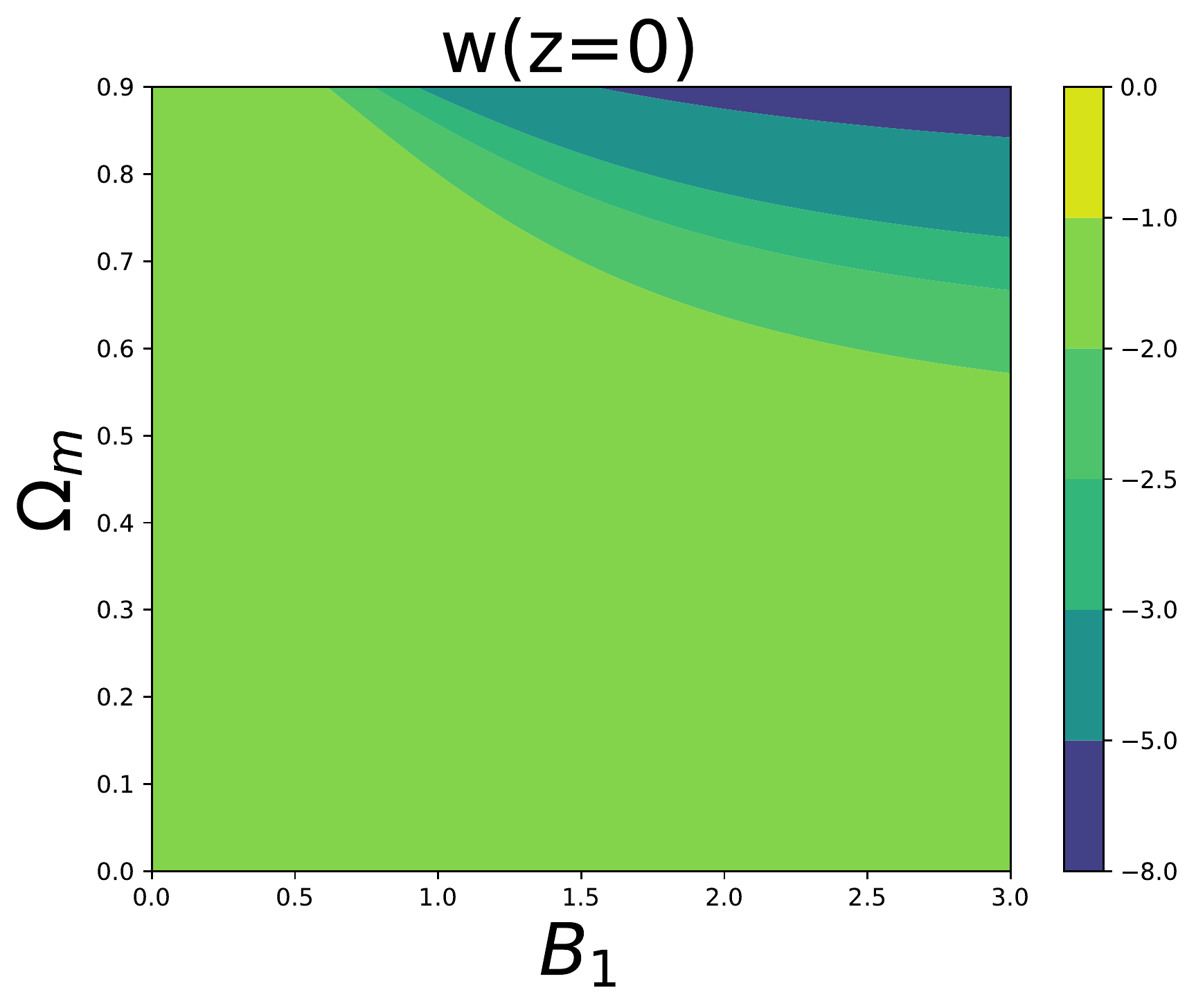}
  \captionof{figure}{Present day $w$ as a function of $B_1$ and $\Omega_{m0}$.} 
  \label{fig:wde_contour}
  \end{minipage}
  \end{minipage}



\section{Growth of Perturbations and Large Scale Structure}
\label{sec:grow_lss}
Next we study the effects of Bimetric gravity on the growth of matter perturbations. We study linear perturbations as well as second-order perturbations. Second-order perturbations affect the skewness ($S_3$) of the matter density field. We study the effects of modified theory parameters on skewness. We use the formalism of Multamaki et al. \cite{10.1046/j.1365-8711.2003.06880.x} to study the growth of matter perturbations but one can also use the formalism for modified gravity models by Lue et. al. \cite{2004PhRvD..69d4005L} and both of them give similar results.

\subsection{Formalism \& Equations}
\label{subsec:formalism_equations}
We start with the formalism of Multamaki et al. \cite{10.1046/j.1365-8711.2003.06880.x}, wherein Raychaudhuri's equation is used to derive a general equation for the growth of perturbations at large scales. We should point out that Bimetric gravity is a modified theory of gravity containing only pressure-less matter in the energy budget of the Universe at late times (one can ignore the contribution from radiation at late times). The formalism developed by Multamaki et al. \cite{10.1046/j.1365-8711.2003.06880.x} to study the matter density perturbations for such modified theory of gravity is briefly discussed below.

\vspace{5mm}
\noindent
The Raychaudhuri's equation for a shearless and irrotational field $v^\mu$ is given by
\begin{equation}
    \dot{\Theta} + \frac{\Theta^2}{3} = R_{\mu\nu}v^\mu v^\nu,
\end{equation}
where
\begin{equation}
    \Theta = \nabla_\mu v^\mu, \qquad \theta = \nabla_i v^i.
\end{equation}
As shown in  \cite{10.1046/j.1365-8711.2003.06880.x}, this can be related to average Hubble expansion rate ($\bar{H}$) and locally perturbed expansion rate $H$ as
\begin{equation}
    \frac{\dot{\theta}}{a}+\frac{\theta}{a}\bar{H}+\frac{\theta^2}{3a^2}
 = 3(\dot{H}+H^2-\dot{\bar{H}}-{\bar{H}}^2).   \end{equation}
Combining the above equation with the continuity equation for pressure-less matter 
\begin{equation} 
\frac{\partial\delta}{\partial t} + (1+\delta)\theta = 0,
\end{equation}
where $\delta$ is the matter overdensity. We get the evolution equation for $\delta$ \cite{10.1046/j.1365-8711.2003.06880.x} as
\begin{equation}
    \frac{d^2\delta}{d\eta^2}+\left( 2 + \frac{\dot{\bar{H}}}{\bar{H}^2} \right)\frac{d\delta}{d\eta} - \frac{4}{3}\frac{1}{1+\delta}\left(\frac{d\delta}{d\eta}  \right)^2 = -3\frac{1+\delta}{\bar{H}^2}\left[\left( \dot{H} + H^2 \right) -\left( \dot{\bar{H}} + \bar{H}^2 \right)  \right], 
    \label{pert_eqn_1}
\end{equation}
where overdot represents derivative w.r.t. time and $\eta\equiv \ln(a) $. Quantities with overbar represent background quantities.
Following \cite{10.1046/j.1365-8711.2003.06880.x}, we can expand the r.h.s of equation (\ref{pert_eqn_1}) as
\begin{equation}
    3\frac{1+\delta}{\bar{H}^2}\left[\left( \dot{H} + H^2 \right) -\left( \dot{\bar{H}} + \bar{H}^2 \right)  \right] 
    = 3(1+\delta)\sum_{n=1} c_n \delta^n.
\end{equation}
We further expand $\delta$ as \cite{10.1046/j.1365-8711.2003.06880.x}
\begin{equation}
    \delta = \sum_{i=1}^\infty \frac{D_i(\eta)}{i!}\delta_0^i,
\end{equation}
where $\delta_0$ is the small perturbation. With this we get the linear and second-order perturbation equations as \cite{10.1046/j.1365-8711.2003.06880.x}

\begin{equation}
    D_1^{\prime\prime} + \left(2+\frac{\dot{\bar{H}}}{\bar{H}^2} \right)D_1^\prime
    +3c_1D_1 = 0,
\end{equation}
and
\begin{equation}
    D_2^{\prime\prime} + \left(2+\frac{\dot{\bar{H}}}{\bar{H}^2} \right)D_2^\prime
    -\frac{8}{3}{D_1^\prime}^2+ 3c_1D_2 + 6(c_1+c_2)D_1^2 = 0.
\end{equation}
For Bimetric gravity, which we are considering here, we get
\begin{equation}
c_1 = \frac{\frac{-2\Omega_m}{a^3}\left(\frac{\Omega_m}{2a^3}+\frac{B_0}{6}\right)\left(\left(\frac{\Omega_m}{2a^3}+\frac{B_0}{6}\right)^2+\frac{B_1^2}{3}\right)-\frac{2\Omega_m}{a^3}\left(\left(\frac{\Omega_m}{2a^3}+\frac{B_0}{6}\right)^2+\frac{B_1^2}{3}\right)^{3/2}-\frac{\Omega_m^2B_1^2}{a^6}}{8\left(\frac{\Omega_m}{2a^3}+\frac{B_0}{6}+\sqrt{\left(\frac{\Omega_m}{2a^3}+\frac{B_0}{6}\right)^2+\frac{B_1^2}{3}}\right)\left(\left(\frac{\Omega_m}{2a^3}+\frac{B_0}{6}\right)^2+\frac{B_1^2}{3}\right)^{3/2}},
\end{equation}
and 
\begin{equation}
c_2 = \frac{\frac{-8\Omega_m^2 B_1^2}{a^6}\left(\left(\frac{\Omega_m}{2a^3}+\frac{B_0}{6}\right)^2+\frac{B_1^2}{3}\right)+\frac{9\Omega_m^3 B_1^2}{a^9}\left(\frac{\Omega_m}{2a^3}+\frac{B_0}{6}\right) }{96\left(\frac{\Omega_m}{2a^3}+\frac{B_0}{6}+\sqrt{\left(\frac{\Omega_m}{2a^3}+\frac{B_0}{6}\right)^2+\frac{B_1^2}{3}}\right)\left(\left(\frac{\Omega_m}{2a^3}+\frac{B_0}{6}\right)^2+\frac{B_1^2}{3}\right)^{5/2}},
\end{equation}
We solve for linear and second-order perturbations for values of $B_1$ and $\Omega_{m}$. We set the initial conditions at redshift $z = 1000$, assuming that model reproduces the Einstein-de Sitter Universe at that epoch. Solving these equations, the linear growth rate is shown in figure \ref{fig:linear} while the evolution of the second-order perturbations are shown in figure \ref{fig:2nd_o}. As shown in figure \ref{fig:linear}, for smaller valueus of $B_{1}$, the linear growth is similar to $\Lambda$CDM model. But for higher values, in particular for values $B_{1} \geq 1.4$, there is an increase in growth around $z \sim 1$. This can be possibly due to some extra attractive gravitational pull provided by the Bimetric gravity for such values of $B_{1}$.

For the second order perturbation as shown in figure \ref{fig:2nd_o}, we get the similar behaviour where the deviation from $\Lambda$CDM increases for larger values of $B_{1}$.

\begin{minipage}{0.975\textwidth}
\centering
 
  \begin{minipage}[b]{0.48\textwidth}
    \centering
   \includegraphics[width=\textwidth]{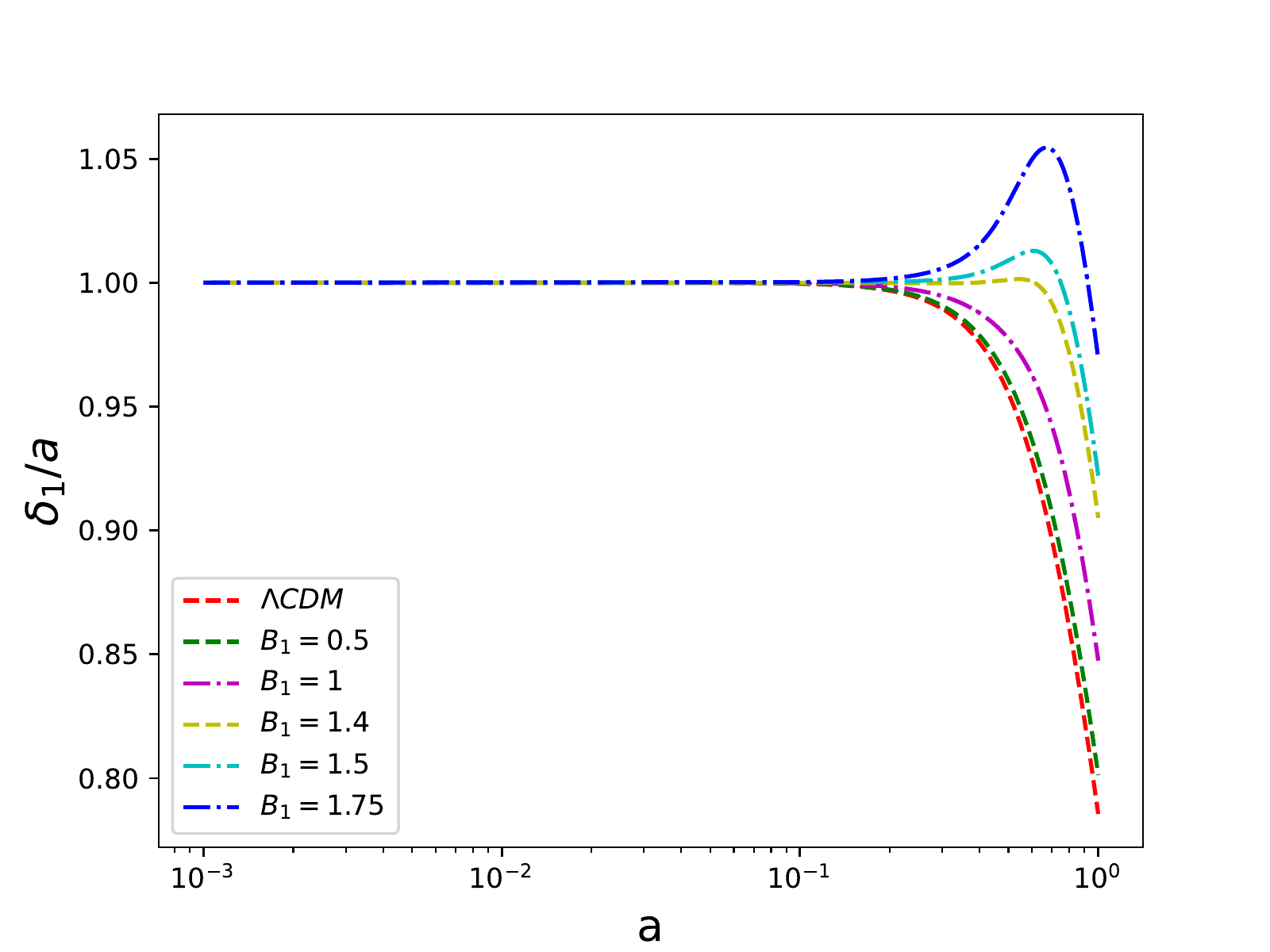}
  \captionof{figure}{Linear perturbations for different values of parameter $B_1$. For higher values of $B_1(>1.4)$, there is a very distinct feature of a brief epoch with growth faster than $\Lambda CDM$ as well as Einstein-diSitter.}
  \label{fig:linear}
    \end{minipage}
     \hfill
     \begin{minipage}[b]{0.48\textwidth}
    \centering
     \includegraphics[width=\textwidth]{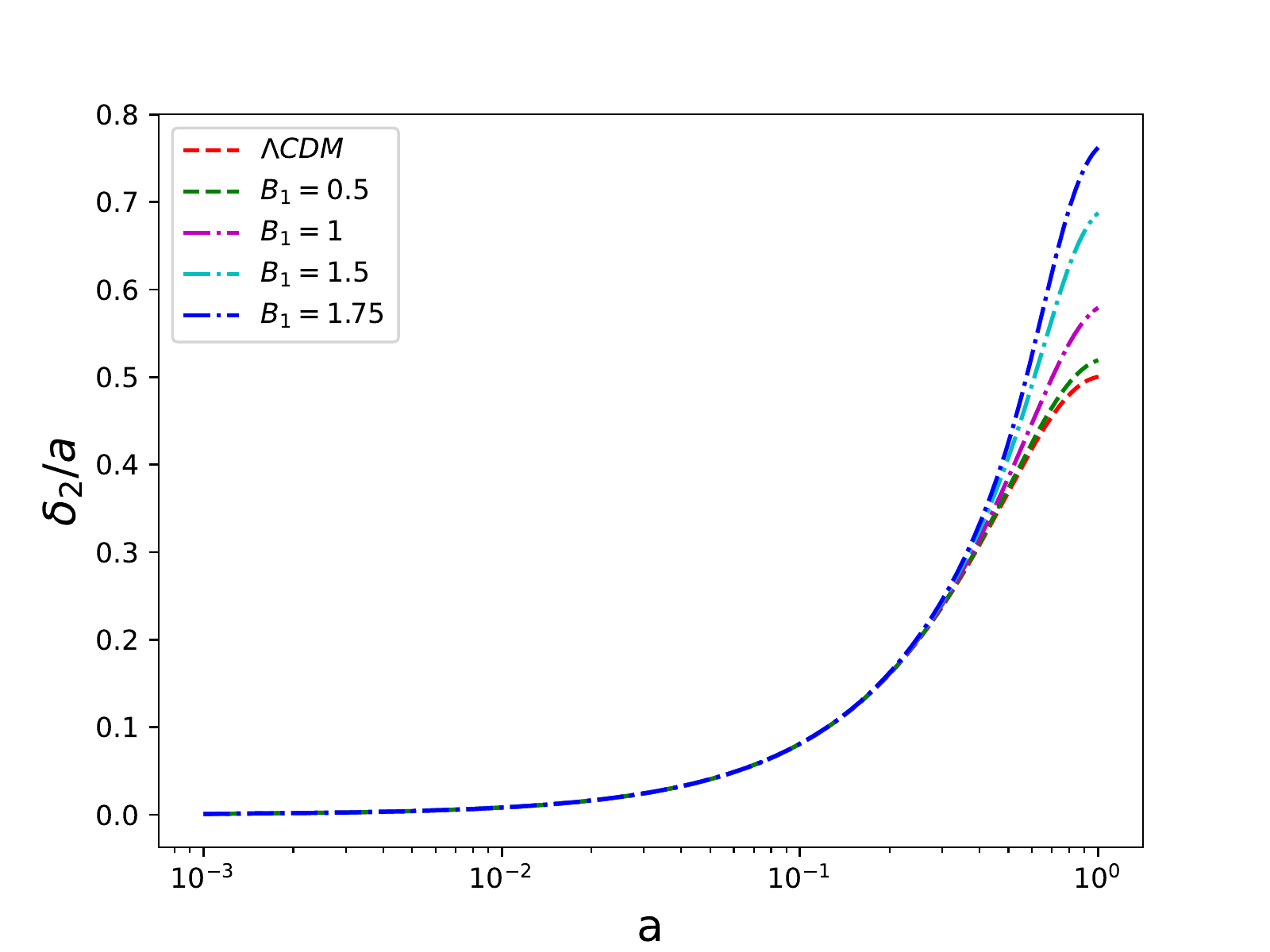}
  \captionof{figure}{Second order perturbations for different values of parameter $B_1$. }
  \label{fig:2nd_o}
  \end{minipage}
  \end{minipage}

\subsection{Observables}
\label{subsec:observable}
\subsubsection{$f\sigma_8$}
Linear theory calculations can be used to predict the growth and clustering of structures at appropriate length scales and times.  Redshift surveys provide an estimate of a combination of linear growth $\delta$, its derivative given by growth factor $f$ and its rms fluctuation at the length scale of $8h^{-1}~Mpc$ given by the parameter $\sigma_8$. The combination $f\sigma_8$ \cite{2009JCAP...10..004S} is an important observable for perturbed Universe at linear scale. Here the growth factor $f$ and the $\sigma_{8}$ parameters are given as
\begin{equation}
    f(a)  = \frac{d (log(\delta))}{d (log (a))},
\end{equation}
and
\begin{equation}
    \sigma_8(a) = \sigma_8(a=1)\frac{\delta(a)}{\delta(1)}.
\end{equation}
In figure \ref{fig:fs8}, we show the $f\sigma_8$ for Bimetric gravity for the different values of the parameter $B_{1}$ along with the $\Lambda$CDM model. We observe that we always get larger $f\sigma_8$ for larger values of $B_{1}$ compared to $\Lambda$CDM ($B_{1}=0$). Moreover for larger values of $B_{1}$, the $f\sigma_{8}$ behaviour for Bimetric gravity models are not consistent in the redshift range $z \sim 0.25-0.5$. In this plot, we fix the value of $\sigma_{8}$ at $z=0$ by Planck-2018 observations \cite{planck_2018_cosmo} assuming a $\Lambda$CDM model. Hence for this figure, one expects that for higher values of $B_{1}$, a lower value for $\sigma_{8}$ may be necessary to make the $f\sigma_{8}$ for Bimetric theory more consistent with the observational data.

\begin{minipage}{0.975\textwidth}
\centering
 
  \begin{minipage}[b]{0.48\textwidth}
    \centering
   \includegraphics[width=\linewidth]{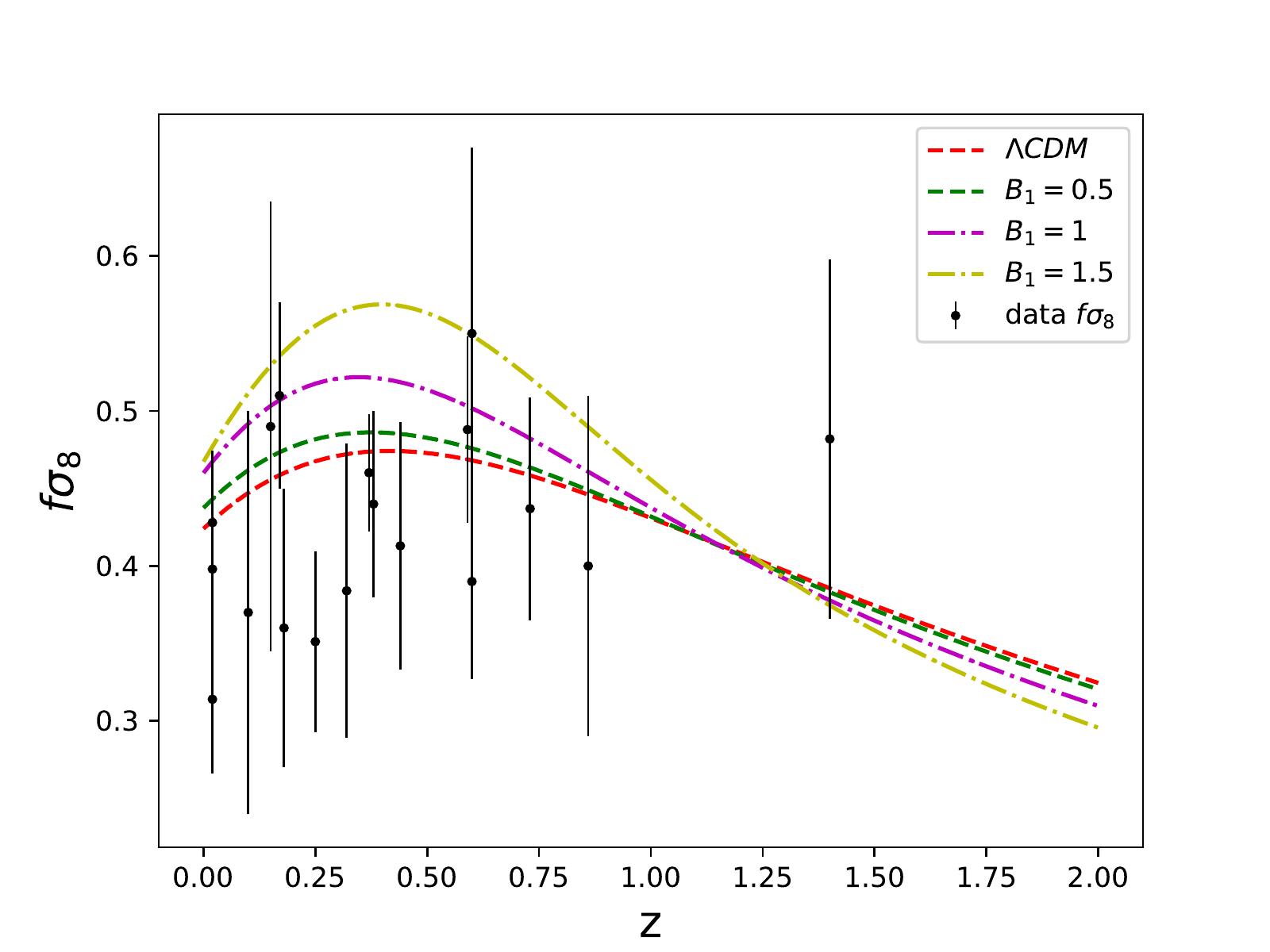}
 \captionof{figure}{Combination $f\sigma_8$ as a function of redshift $z$ for different theories.  We also plot the data points from observations \cite{Nesseris_2017}. }
  \label{fig:fs8}
    \end{minipage}
     \hfill
     \begin{minipage}[b]{0.48\textwidth}
    \centering
    \includegraphics[width=\linewidth]{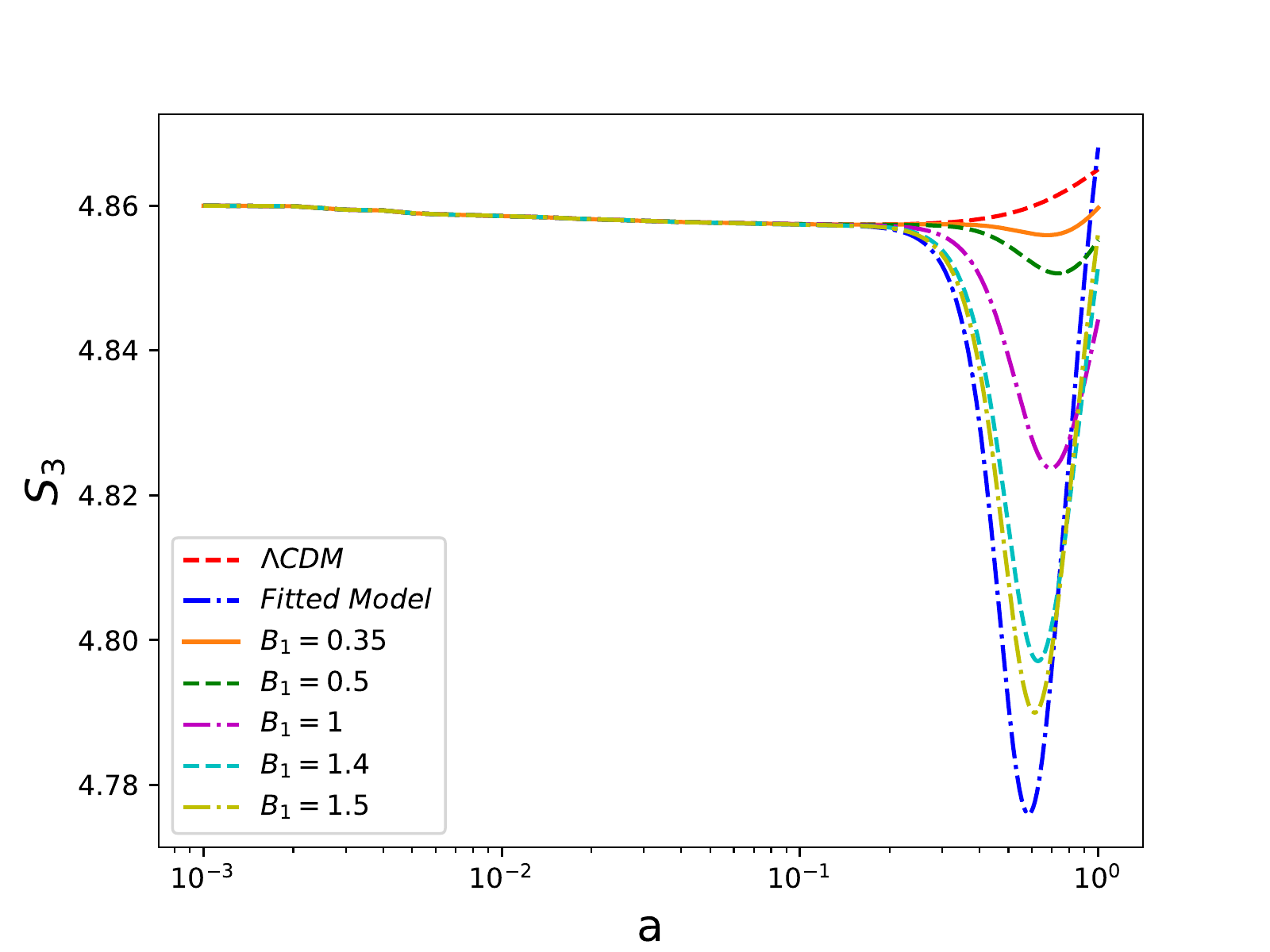}
 \captionof{figure}{Skewness $S_3$ as a function of scale factor for different values of $B_1$.}
  \label{fig:S3}
  \end{minipage}
  \end{minipage}

\begin{figure}[h!]
  \includegraphics[width=0.5\linewidth]{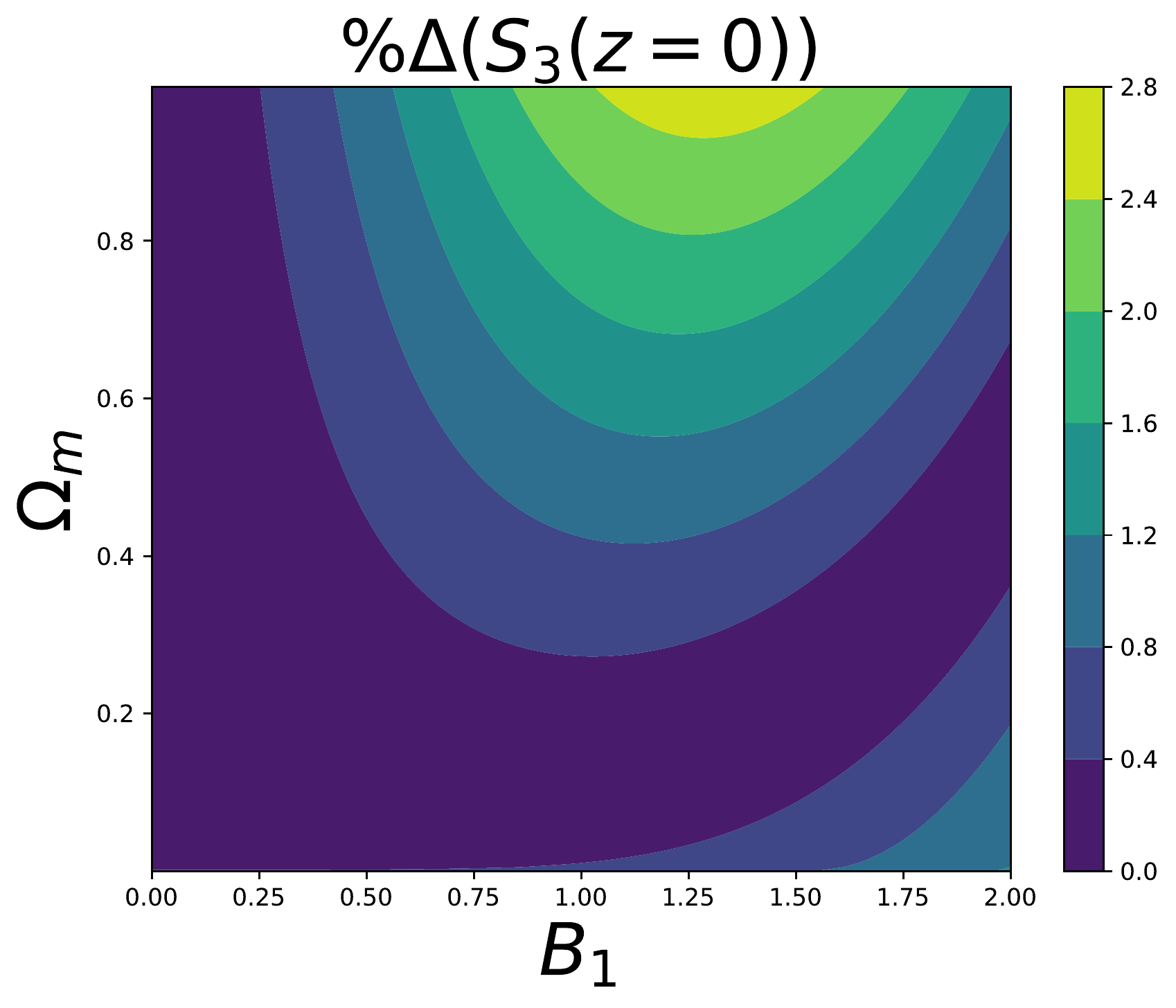}
  \caption{Deviation of $S_3$ from $\Lambda CDM$ model for different values of $B_1$. While the linear growth rate and second-order perturbations show huge differences, the percentage difference in the combination probed by $S_3$ is a few percent.}
  \label{fig:S3_cont}
\end{figure}

\subsubsection{Skewness (S3)}
\label{subsubsec:s3}
Second-order perturbations provide further connection with statistics of observed perturbations. Gaussian initial conditions evolve into non-gaussian distribution with time and the extent of non-gaussianity depends on dynamics of the individual fluid components of the universe or the theory of gravity which evolves the whole system. Mode coupling leads to an imbalance in the distribution of voids and overdense regions \cite{Bernardeau_2002}. Second-order perturbations play a role in this and can be related to the skewness of the density field as \cite{10.1046/j.1365-8711.2003.06880.x,Bernardeau_2002}
\begin{equation}
    S_3 = \frac{\langle\delta^3 \rangle}{\langle\delta^2 \rangle^2},
\end{equation}
which can be written as
\begin{equation}
    S_3 = 3\frac{D_2}{D_1^2}.
\end{equation}
$S_3$ can be sensitive to underlying dark energy characteristics \cite{2022arXiv220208355E} or modification of GR.
Here we show that the growth of perturbations is sensitive to parameter $B_1$ as we illustrate in figures 3 and 4.  In figure \ref{fig:S3}, we show the evolution of $S_3$ for different values of $B_1$. In figure \ref{fig:S3_cont} we show  the present day percentage difference of $S_3$ from $\Lambda$CDM model as a function of $B_{1}$ and $\Omega_{m0}$. This can be used to distinguish the Bimetric models from the $\Lambda$CDM using the $S_{3}$ measurements in near future.

\section{Observational Constraints on Bimetric Model}
\label{subsubsec:DataA}

Given the behaviour of Bimetric gravity in terms of different observables, we now study the observational constraints on Bimetric gravity using low redshift cosmological observations. In this regard, we do the Markov Chain Monte Carlo (MCMC) analysis using different latest cosmological observational data to put constraints on the model parameters for the Bimetric gravity. The analysis is performed using the EMCEE hammer \cite{Foreman_Mackey_2013}, a PYTHON implementation of the MCMC sampler.

\noindent
We use the following data:

\begin{itemize}
\item
Pantheon+ and SH0ES data \cite{Brout:2022vxf} ;
\item 
The Gold-2017 set for the $f\sigma_8$ data \cite{Nesseris_2017,Macaulay:2013swa};
\item 
The Baryon acoustic oscillations (BAO) measurements by the completed Sloan Digital Sky Survey (SDSS) lineage of experiments on large scales \cite{eBOSS:2020yzd};
\item
The angular diameter distances measured using water megamasers under the Megamaser Cosmology Project \cite{Reid:2008nm}.
\end{itemize}

\begin{table*}[h!]
\centering
\begin{tabular}{|c|l|}
\hline
Parameter&Prior\\
\hline\hline

$\Omega_m$  & $[0.0,0.9]$ \\
\hline
$h$&$[0.6,0.8]$\\
\hline
 $B_1$    &  $[0,5]$\\ 
 \hline
 $r_d$  &  $ [130,160]$   \\
 \hline
 $\sigma_8$  &  $[0.01,0.9]$   \\
  \hline
 $M$  &  $[-19.40,-19.00]$  \\
\hline
\end{tabular}
\caption{The range of the uniform priors for the parameters used for the MCMC analysis.}
\label{tab1}
\end{table*}

\begin{figure}[H]
  \centering
 \includegraphics[width=0.78\textwidth]{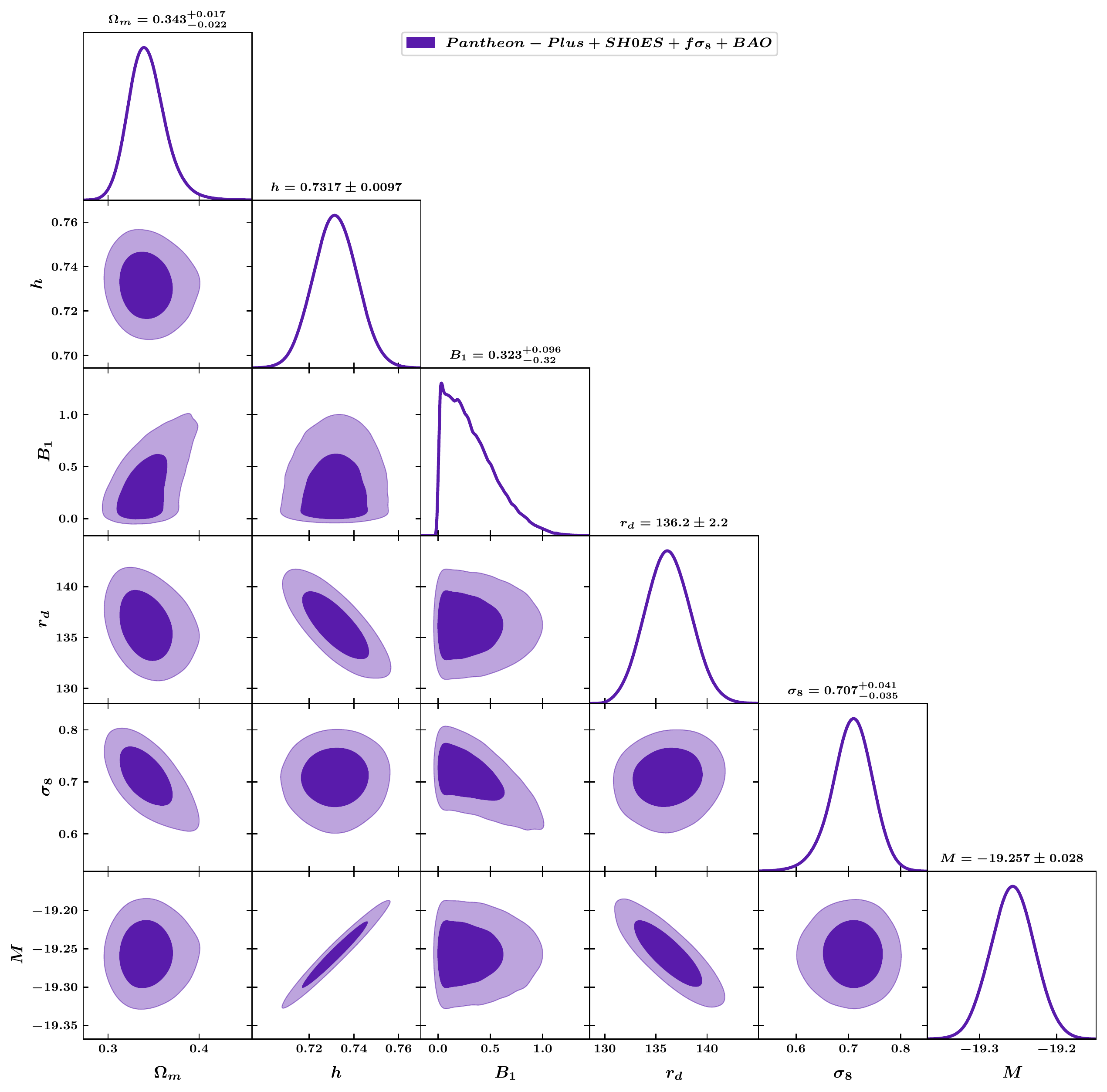}
  \caption{Marginalized posterior distribution of the set of parameters $\Omega_m$, $h$, $B_1$, $r_d$, $\sigma_8$ and $M$ and their corresponding 2D confidence contours, obtained from the MCMC analysis for the Bimetric gravity utilizing all
the data sets mentioned in section \ref{subsubsec:DataA}}
  \label{fig:DA}
\end{figure}

We use the uniform priors given in Table \ref{tab1} for the model parameters for the Bimetric gravity. The posterior probability distributions and their corresponding confidence contours for different parameters are shown in figure (\ref{fig:DA}). As one can see from this figure, a substantial deviation from $\Lambda$CDM model in terms of the paramerer $B_{1}$ is allowed by the data although the $\Lambda$CDM behaviour ($B_{1}=0$) is still allowed. The reconstructed effective dark energy equation of state and the reconstructed $S_{3}$ parameters as a function of redshifts are shown in figures (\ref{fig:eos_bf}) and (\ref{fig:S3_recon}). The constrained Bimetric model gives a phantom-like effective dark energy equation of state. The data also allows substantial deviation in the parameter $S_{3}$ from $\Lambda$CDM model behaviour although the $\Lambda$CDM behavious for $S_{3}$ is also allowed in the constrained behaviour of $S_{3}$.

\begin{minipage}{0.89\textwidth}
\centering
 
  \begin{minipage}[b]{0.51\textwidth}
    \centering
    \includegraphics[width=\textwidth]{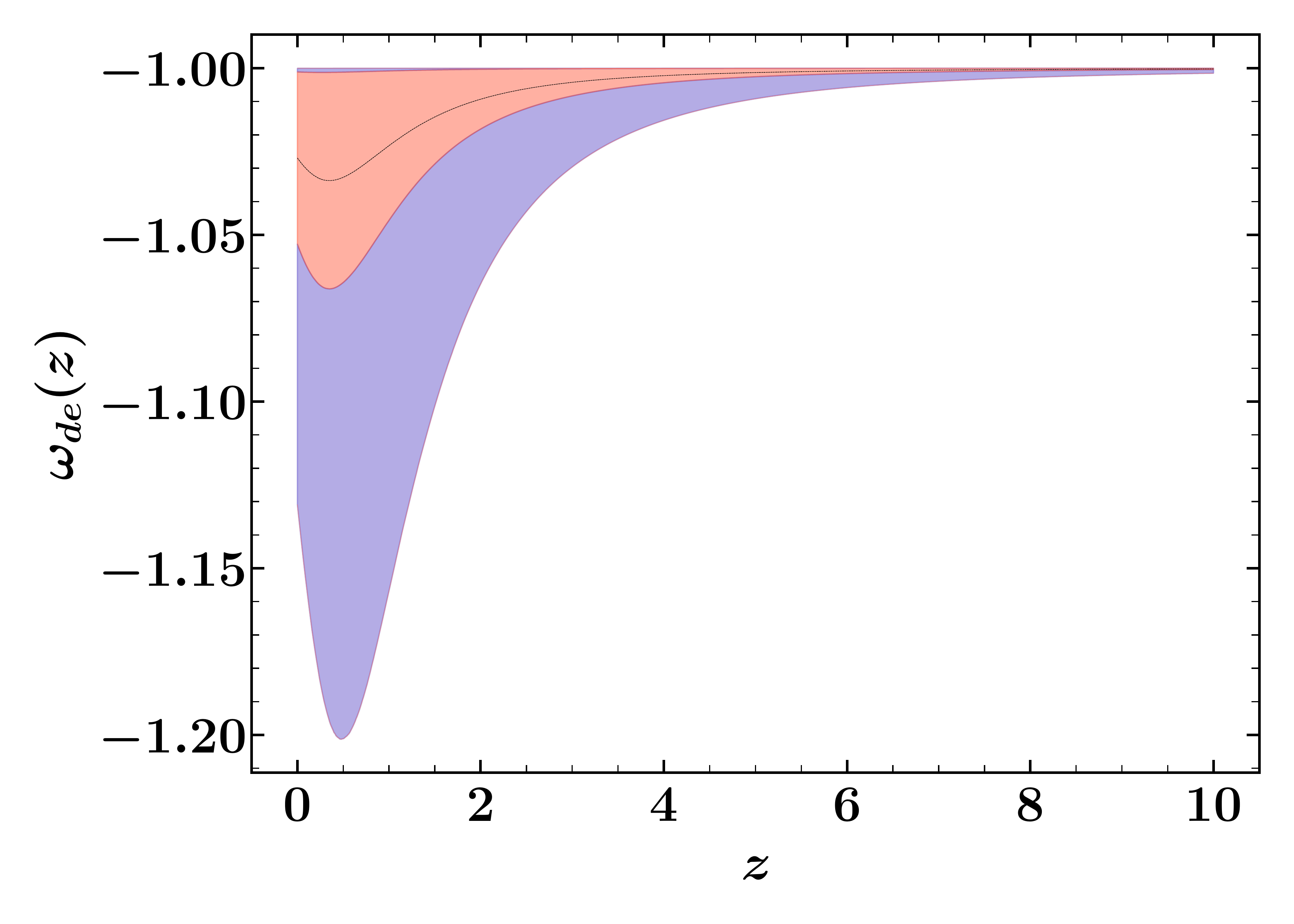}
  \captionof{figure}{Reconstructed equation of state $\omega_{de}$ as a function of redshift z. Black line is the best fit value with shaded regions as $1\sigma$ and $2\sigma$ for the inner and the outer shaded region respectively.}
  \label{fig:eos_bf}

    \end{minipage}
     \hfill
     \begin{minipage}[b]{0.47\textwidth}
    \centering
    \includegraphics[width=\textwidth]{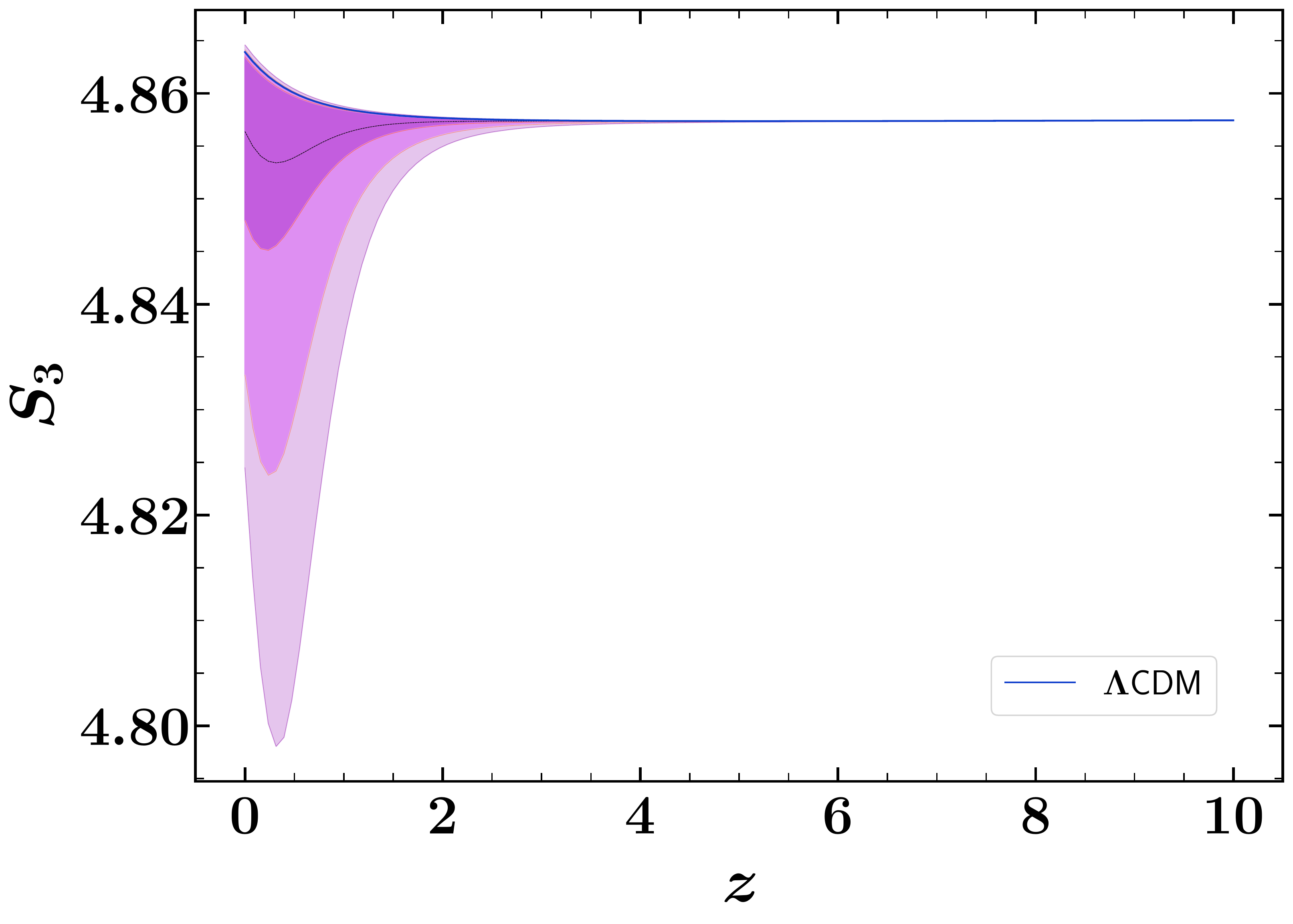}
    \captionof{figure}{Reconstructed skewness $S_3$ as a function of redshift z. Black line is the best fit value with shaded regions as $1\sigma$, $2\sigma$ and $3\sigma$ for the innermost to the outermost shaded region. Blue line is for the $\Lambda$CDM model.}
    \label{fig:S3_recon}
  \end{minipage}
  \end{minipage}

\section{Integrated Sachs Wolfe (ISW) effect}
\label{subsubsec:sw}
Given the observational constraints on the Bimetric gravity considered in this analysis, we study how far the ISW signal in this model deviates from the $\Lambda$CDM model within that constraints. CMB photons traveling through evolving spacetime, traverse  potentials created by matter inhomogeneity and undergo changes in their wavelengths. This contributes to anisotropies of CMB spectrum and is dubbed ISW effect \cite{1967ApJ...147...73S,2014PTEP.2014fB110N}. The effect can be detected by cross-correlation of Large Scale Structure (LSS) tracers with CMB anisotropies \cite{2014PTEP.2014fB110N,2004PhRvD..69d4005L,1996PhRvL..76..575C,2022JCAP...04..033K} which can be used as a probe for theories giving dark energy effects. Here we follow the formalism of Lue et al. \cite{2004PhRvD..69d4005L} who gave a general prescription for studying the ISW effect in modified gravity theories. We calculate the evolution of time derivatives of potentials for our Bimetric model and compare it with the standard $\Lambda$CDM model.

Follwing the prescription by Lue et al. \cite{2004PhRvD..69d4005L}, we characterize background expansion in Bimetric gravity theory by a function $g(x)$ as
    \begin{equation}
        g(x) = \left(\frac{H}{H_0}\right)^2,
    \end{equation}
    with $x$ defined as 
    \begin{equation}
        x \equiv \frac{8\pi G\rho_m}{3H_0^2}.
    \end{equation}
    For example, in $\Lambda CDM$, the function takes the form
    \begin{equation}
        g(x) = x + \Omega_{\Lambda}.
    \end{equation}
    For the Bimetric gravity, $g(x)$ is given by
    
    \begin{equation}
        g(x) = \frac{1}{2}x + \frac{B_0}{6} + \sqrt{\left( \frac{1}{2}x + \frac{B_0}{6} \right)^2 + \frac{B_1^2}{3}   }.
    \end{equation}

    \noindent
We start with the following convention for the perturbed metric \cite{2004PhRvD..69d4005L}

    \begin{equation}
        ds^2 = -(1+2\Psi)dt^2 + a^2(1+2\Phi)(dr^2+r^2d\Omega^2).
    \end{equation}

\noindent
The ISW effect is proportional to a combination of temporal derivatives of potentials as
    
    \begin{equation}
        A(\dot{\Psi}-\dot{\Phi})  = \left[(1-f^{MG})(g^\prime +(3/2) g^{\prime\prime}x) +\frac{3}{2}(5xg^{\prime\prime}+3x^2g^{\prime\prime\prime})  \right]D_+ ,
    \end{equation}

    \noindent
wherein $D_+$ is linear first order perturbation and $f$ is growth rate defined as $\frac{d(lnD_+)}{d(ln a)}$. 
In figure \ref{fig:isw_potn}, we compare this term for our constrained Bimetric gravity (for best fitted parameter values) with $\Lambda$CDM model as constrained by Planck-2018 . We see that the observationally constrained Bimetric model does not differ much from the $\Lambda$CDM model.\vspace{5mm}

\noindent
Finally, the cross-correlated ISW signal ($w_{gT}$) between LSS and CMB can be calculated as 

    \begin{eqnarray}
        w_{gT} = 3T_0\Omega_{m0}b(2\pi)^2\frac{H_0}{c^3}\int dz\sqrt(g) && D_+^2\left[(1-f)(g^\prime +(3/2) g^{\prime\prime}x) +\frac{3}{2}(5xg^{\prime\prime}+3x^2g^{\prime\prime\prime})  \right]\nonumber \\&& w_g(z) \int\frac{dk}{k}P(k)J_0(k\theta\chi).
    \end{eqnarray}
\begin{minipage}{0.975\textwidth}
\centering
 
  \begin{minipage}[b]{0.48\textwidth}
    \centering
   \includegraphics[width=\linewidth]{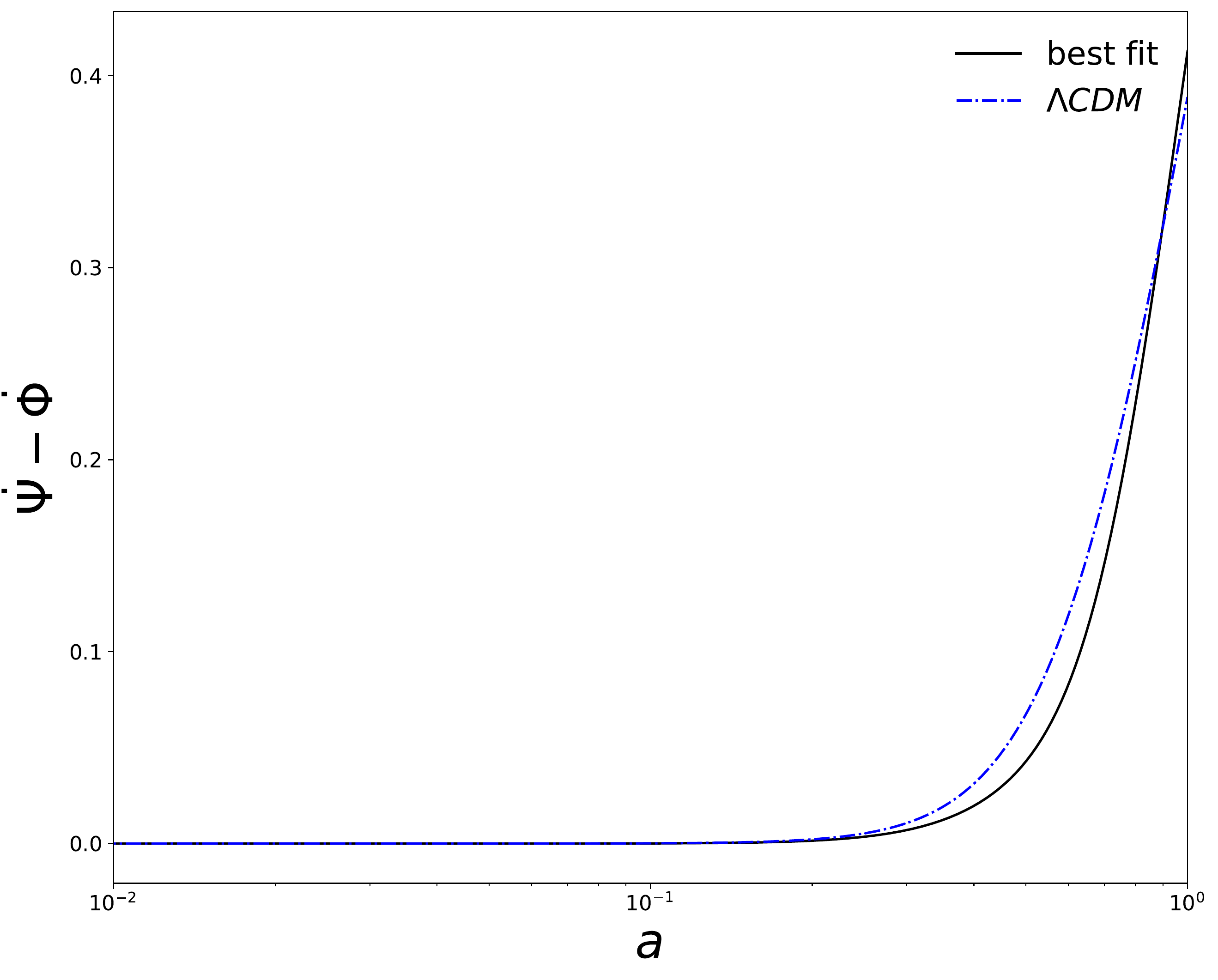}
  \captionof{figure}{Comparison of the evolution of temporal derivatives of the potentials. Differences in linear growth rate are translated here as well.}
  \label{fig:isw_potn}
    \end{minipage}
     \hfill
     \begin{minipage}[b]{0.48\textwidth}
    \centering
    \includegraphics[width=\linewidth]{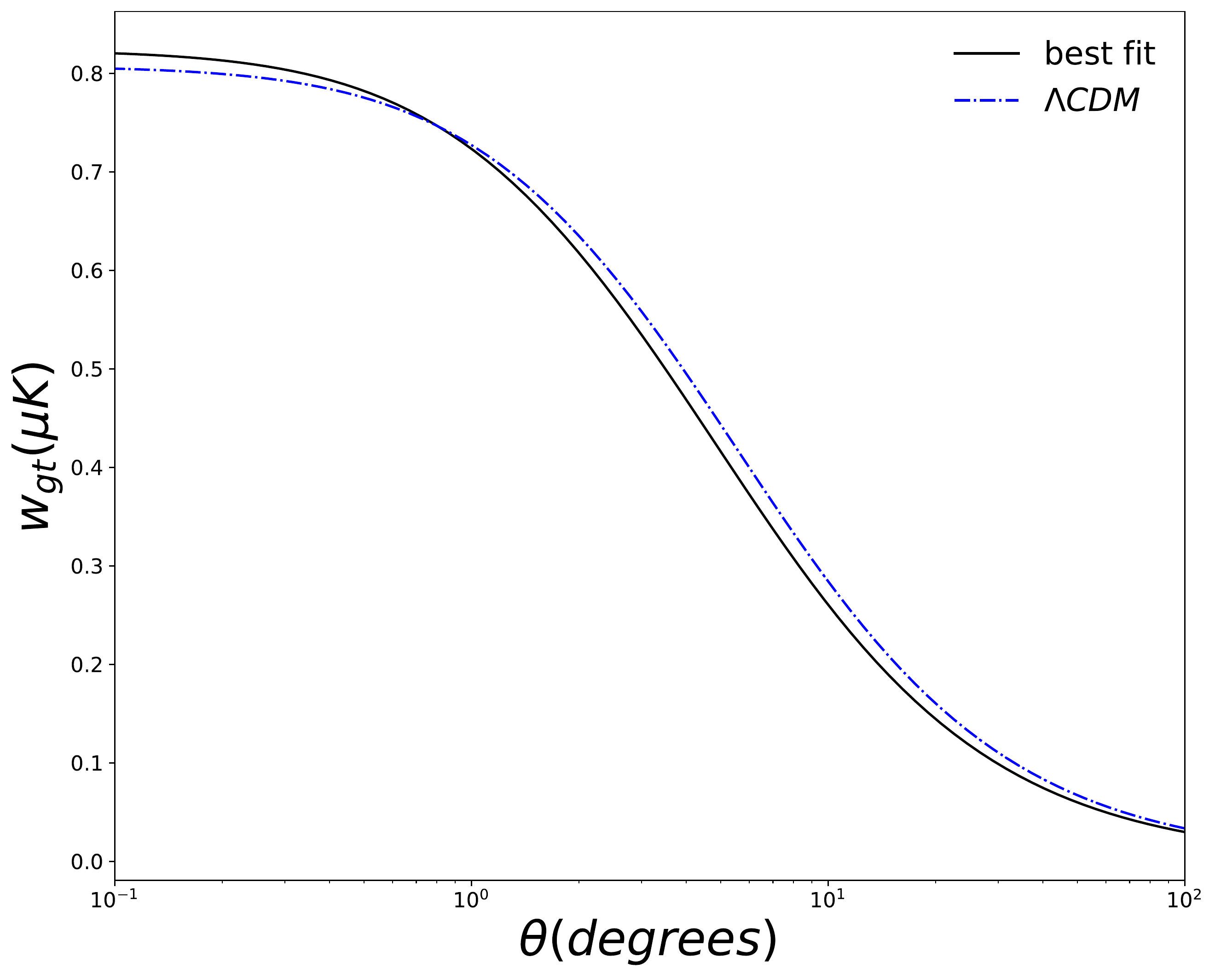}
    \captionof{figure}{Galaxy Temperature correlation for the best fit Bimetric model along with $\Lambda CDM$. The two are very similar.}
  \label{fig:isw_wgt}
  \end{minipage}
  \end{minipage}

\vspace{7mm}

Here $T_0$ is the present CMB temperature, $b$ is the bias factor (assumed constant here), $P(k)$ is present-day matter power spectrum, $\chi$ is comoving distance as a function of $z$, $J_0$ is zeroth Bessel function and $w_g(z)$ is the survey-dependent galaxy selection function. For $T_0$, we take the value $ 2.725 \mu K$, $b$ is taken from Lue et al. \cite{2004PhRvD..69d4005L} that is $5.47$. We use the $w_g(z)$ of Takada \& Jain \cite{2009MNRAS.395.2065T} with mean redshift of $0.49$.\\
This correlation function for best fitted Bimetric gravity as well as for $\Lambda$CDM model are plotted in figure \ref{fig:isw_wgt} and it is evident that the cross-correlated ISW signal in Bimetric gravity is similar to $\Lambda$CDM model.



\section{Conclusions}
\label{sec:summary_conclusion}
We study the cosmological evolution in a subclass of Bimetric gravity model, where only the parameters $B_{0}$ and $B_{1}$ are nonzero. We show that the effective dark energy behaviour in such a modified gravity theory can be phantom-like for a large range of parameter values. Moreover the model admits a cosmological constant that can be positive or negative depending on the values of the parameter $B_{1}$ and $\Omega_{m}$. This can naturally mimic a cosmological evolution where the Universe contains a phantom like dark energy plus a negative Cosmological Constant apart from the standard matter. As shown recently such set up can be useful to solve the Hubble Tension \cite{2022MNRAS.tmp.2681S}.

We also study the linear and second order growth of matter fluctuations in the Bimetric gravity. We find that the growth of both linear and second order perturbations are strongly dependent on the values of parameter $B_1$ that signifies the deviation from the corresponding $\Lambda CDM$ limit. This results in significant deviations of observables like "$f\sigma_{8}$" and "Skewness" parameter $S_{3}$ from the $\Lambda$CDM behaviour for higher values of the parameter $B_{1}$. 

With these observations, we subsequently constrain the Bimetric model with low-redshift observational data from SnIa Observation ( Pantheon+ and SH0ES), BAO observations as well as Growth measurements. It shows that the data allow significant deviation from $\Lambda$CDM behavior although $\Lambda$CDM limit of Bimetric theory ($B_{1} =0$) is also consistent with the data.

Finally we calculate the ISW signal by cross-correlating the CMB and LSS signals for our best fit Bimetric gravity model and show that it is mostly similar to the $\Lambda$CDM model as constrained by Planck-2018.

To conclude, we show that the low-redshift observations allow Bimetric gravity that behaves differently than $\Lambda$CDM. This motivates us to study the behaviour of CMB fluctuations in such models and see whether they are consistent with the Planck-2018 measurements. We plan to study this in the near future. 

\section*{Acknowledgements}

 AAS acknowledges the funding from SERB, Govt of India under the research grant MTR/20l9/000599.  MPR acknowledges the funding from SERB, Govt of India under the research grant no: CRG/2020/004347. SAA is funded by UGC non-NET Fellowship scheme. The authors also acknowledge the use of High Performance Computing facility Pegasus at IUCAA, Pune, India.

\clearpage
\bibliography{ref}

\bibliographystyle{ieeetr}

\end{document}